\documentclass[fleqn,twoside]{article}
\usepackage{espcrc2}

\usepackage{graphicx}
\usepackage[figuresright]{rotating}

\def\be7pg{$^7Be(p,\gamma)^8B$}
\def\xbe7{$^7Be$}
\def\b8{$^8B$}
\def\S17{$S_{17}(0)$}
\def\xs17{$S_{17}$}
\def\s34{$S_{34}(0)$}
\def\xpm{$\pm$}
\newcommand{\AmS}{{\protect\the\textfont2
  A\kern-.1667em\lower.5ex\hbox{M}\kern-.125emS}}

\title{How Accurately Do We Know the Formation of Solar $^8B$?}

\author{Moshe Gai  \address{Laboratory for Nuclear Science at 
Avery Point, University of Connecticut, 1084 Shennecossett Rd., Groton, CT 06340. 
e-mail: moshe.gai@uconn.edu, URL: http://www.phys.uconn.edu}
\thanks{Work Supported by USDOE Grant No. DE-FG02-94ER40870.}}
       
\begin{document}

\begin{abstract}
A detailed examination of current data on \xs17 (as opposed to an 
examination of \S17 only) excludes quoting \S17 with sufficiently 
small uncertainty. In contrast to suggestions that \S17 is now known 
with the accuracy of \xpm 3\%, the exact value of \S17 is dependent on 
the choice of the data and the choice of theory used for extrapolation. 
In addition recent high precision results (including the Seattle data) on 
\xs17 which are in good agreement, still differ on the measured slopes, 
as do theoretical models that predict different d-wave contribution, precluding 
an accurate extrapolation to zero energy of the consistent data. Using a 
common extrapolation of only the consistent high precision data, suggests a 
value of \S17 = 21.2 \xpm \ 0.5 eV-b, but a value equal to or smaller 
than 19.0 eV-b can not be excluded due to the uncertainty in the extrapolation, 
leading to an additional error of $^{+0.0}_{-3.0}$ eV-b. This (unacceptable) situation 
must be cleared by future experiments.
.
\vspace{1pc}
\end{abstract}

% typeset front matter (including abstract)
\maketitle

\includegraphics[width=35pc]{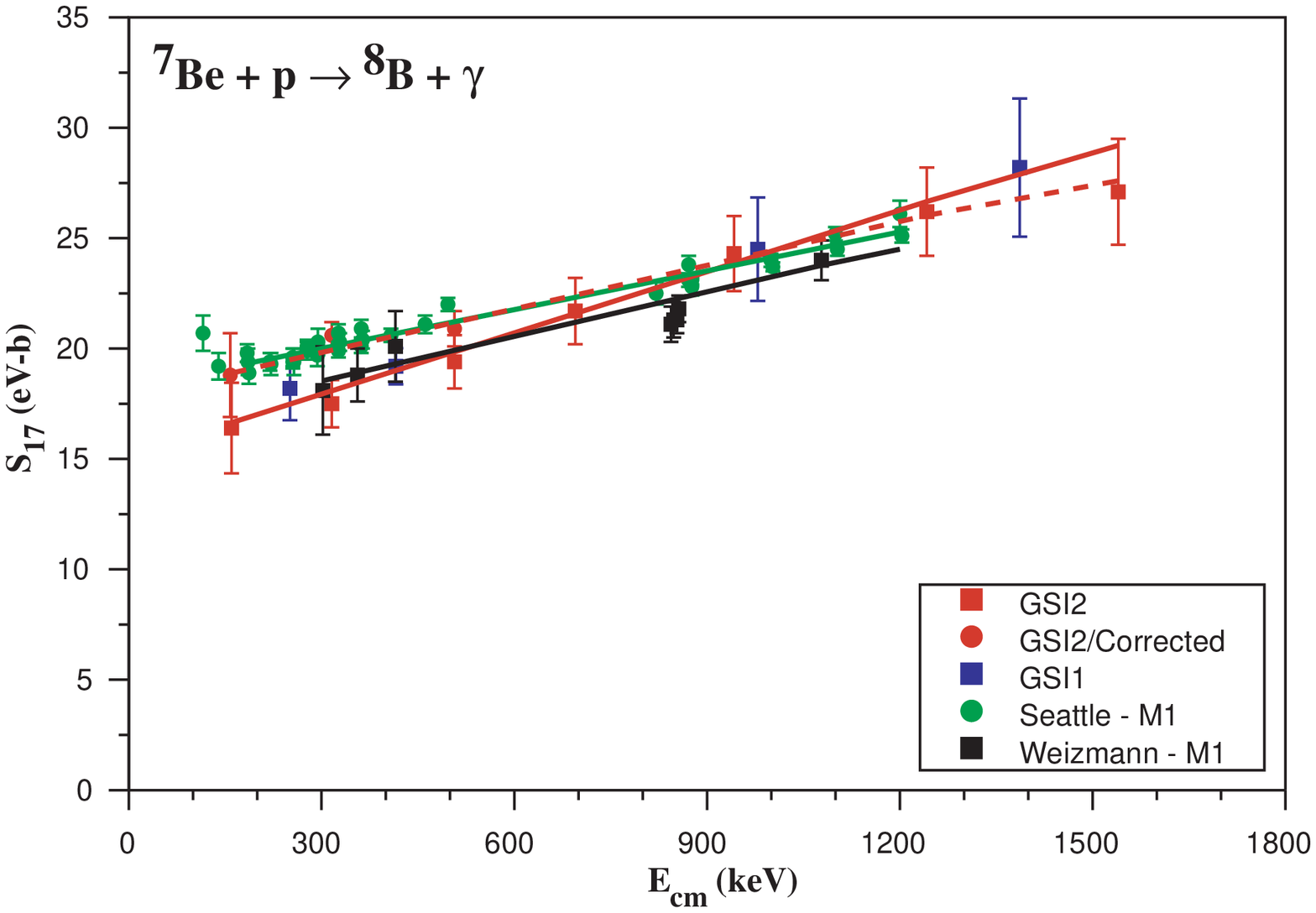}|\\

%\begin{thebibliography}{9}

%\end{thebibliography}

\end{document}